# Structure of System-Specific Hazard Modeling for Physical Infrastructures in a Climate Change Context

Matthieu Dutel

This paper addresses the definition and modeling of hazards in the context of physical systems, particularly infrastructures, that are affected by climate change. The paper highlights that hazards are system-dependent, meaning what constitutes a hazard for one system (e.g., humans) may not be the same for another (e.g., infrastructures). The paper also highlights that extreme intensity distribution may evolve in a region without necessarily constituting a hazard for a specific system. These distinctions are crucial for risk assessment and adaptation strategies in a changing climate. The paper implements a simple threshold-based hazard model - Boolean Hazard Model with Probability - to focus exclusively on system-specific hazard definition in climate-change context. Use cases illustrate how this approach adapts to different systems and hazard types. The model produces time-evolving hazard maps, delivering hazard data tailored to both hazard type and infrastructure characteristics. The model provides hazard data tailored to specific hazard types and physical systems, with a focus on infrastructure. The use cases demonstrate their practical value and potential for broader application in multi-infrastructure and multi-hazard exposure modeling.

## 1. Introduction

**List of Abbreviations**

| **Abbreviation** | **Signification** |
|---|---|
| BHMP | Boolean Hazard Model with Probability |
| HIT | Hazard Intensity Threshold |
| PDR | Physical Damage Rate |

### 1.1. Context and Motivation

The definition of a hazard as an event that can cause damage to a system inherently depends on the system itself. To illustrate this, consider two contrasting systems: a cast-iron pot and the human body. Exposure to 100°C for one hour would likely cause no damage to the pot (physical damage rate near to 0%) because this system is designed to withstand with such a temperature, whereas the same exposure would result in complete damage to the human body (physical damage rate 100% , i.e., death). This analogy highlights that hazard classification depends on system vulnerability. If we define a hazard as the occurrence of potential damage on a system caused by an intensity event. Then, the fact that an intensity constitutes a hazard depends on the system studied. We are in a context of climate change. Climate change is modifying the environmental conditions under which physical systems operate. Global mean temperature is rising, and local effects are complex: the distribution of extreme intensities is shifting in a spatially non-uniform way.

Currently, in France, physical infrastructures operate with high performance levels and support essential activities such as water access, transportation, food supply, lighting, and digital services. However, most infrastructures supporting these operations were designed under the assumption of a stable climate. This assumption is no longer valid. If future extremes exceed design or damage thresholds, physical damage may occur. To anticipate such risks, exposure modeling is used to quantify the number of assets affected by a given hazard. This requires two key inputs: (i) information about the system of interest and (ii) hazard data specific to that system. In this paper, we focus on the second component: system-specific hazard information in a changing climate context.

## 1.2. Framework

System-specific hazard modeling related to climate change relies on two essential inputs: return intensity data and HITs . These two components are inputs for the hazard model, which processes them to generate hazard outputs. Figure 1 illustrates the high-level process linking climate data, system-specific hazard intensity thresholds, and hazard outputs.

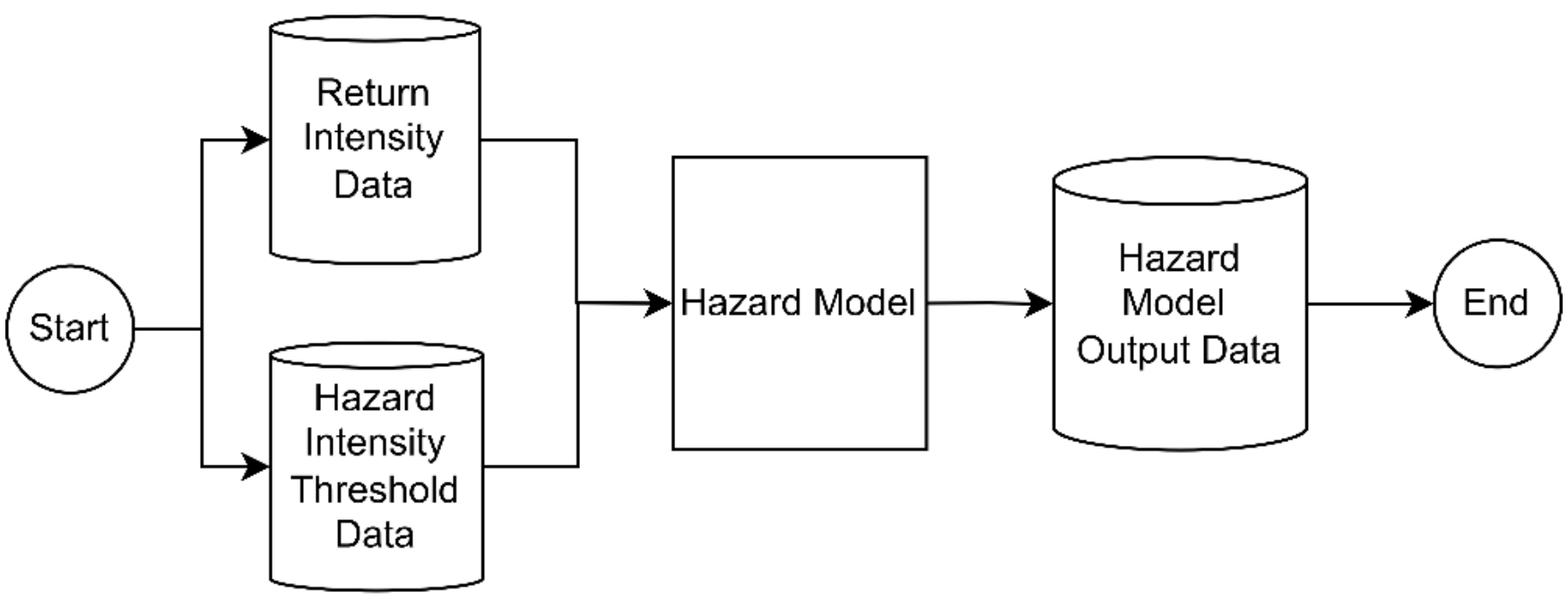

**Figure 1.** High-Level Flowchart of System Specific Hazard Modeling in a Climate Change Context.

Return intensity data represents the estimated values of climate-related intensities for given return periods, obtained from empirical distributions or extreme value statistical models. These distributions can be derived from various sources, such as climate reanalysis datasets, climate projections, or weather station records.

Hazard intensity thresholds establish the link between climate intensity and the characteristics of a specific system. A hazard intensity threshold can be considered as a point of vulnerability information: it represents the intensity level at which a system transitions from experiencing no damage to potentially experiencing damage. In other words, HIT marks the threshold beyond which physical damage becomes possible for the system under consideration. When the return intensity exceeds this threshold, the event is classified as a hazard for the system under study. Climate data alone does not define a hazard specific to a system, rather, a hazard emerges when climate intensity surpasses the system's damage threshold. Beyond the basic HIT, which corresponds to the onset of damage, additional thresholds can be defined for specific physical damage rates. This enables the modeling of hazards to be in relation to different levels of expected damage. Such thresholds are therefore system-specific, climate data-specific, and damage rate specific. This system specific hazard modeling ensures that hazard modeling reflects the vulnerability information point for specific system in a context of changing climate extremes. In this paper, we apply the BHMP framework, which is a subpart of the Boolean Exposure Modeling framework presented in the previous work, Dutel et al. (2025).

### 1.3. Contribution

This paper contributes to system-specific hazard modeling in three ways:

(i) Conceptual: A system-specific definition of climate-related hazards, where hazard classification depends on an explicit system-specific threshold.

(ii) Methodological: Generalization of the BHMP to multiple hazard types (e.g. wind, temperature) and systems (e.g., wind turbines, power towers, asset concerned by a specific Eurocode).

(iii) Practical: Demonstration of HITs derived from vulnerability curves or design standards to generate system-specific hazard maps for exposure modeling.

### 1.4. Structure of the paper.

The following of this paper is organized as follows: Section 2 – Research Gap, explains the location of this paper in the literature. Section 3- Concepts and Definitions, introduces concept and definitions for system-specific hazard modeling in the context of climate change. It defines what constitutes a climate-related hazard for a given system and explains how HITs and return intensity data are combined to classify events as hazards. Section 4 – Models and Methods, presents the hazard model we use: BHMP and details the process for generating hazard maps tailored to different systems, hazard types and source for vulnerability information. Section 5 – Use Cases and Results, demonstrates the application of BHMP to different systems and hazard types. This section presents illustrative use cases using vulnerability curves and design standards, and shows hazard maps for various Physical Damage rates. Section 6- Discussion, interprets the results and emphasizes the importance of system-specific hazard outputs as inputs for exposure quantification. This section compares sources of HIT and discusses limitations and potential improvements. Section 7- Conclusion summarizes contributions of the study.

## 2. Research Gap

Hazard modeling and mapping for a specific system requires hazard information that is specific to the system under consideration.

Climate data has continuously improved over two centuries, from early foundational work (Fourier, 1824 ; Tyndall, 1861 ; Arrhenius, 1897 ; Callendar, 1938), to advances in numerical modeling and climate modeling (Charney et al. 1950; Phillips, 1956; Manabe and Wetherald, 1967; Manabe and Bryan, 1969; Déqué and Piedelievre, 1995; Giorgi, 1990; Cess et al., 1989; Eyring et al., 2016; Meehl et al., 2000), operational platforms (Hersbach et al., 2020; Jacob et al., 2014; Lémond et al., 2011), downscaling (Verfaillie et al., 2017), and more recently machine-learning-based approaches (Kochkov et al., 2024; Lam et al., 2023; Bodnar et al., 2025; Watt-Meyer et al., 2025, 2023). Yet, these data remain climate intensities, independent of the physical characteristics of the systems they may affect. An intensity only becomes a hazard for a specific system when it exceeds a system-specific vulnerability threshold, information that climate data do not contain.

At the same time, vulnerability information exists but is heterogeneous and incomplete : some infrastructures for some climate variable benefit from detailed vulnerability curves (Nirandjan et al., 2024; Reinoso et al., 2020; Scawthorn et al., 2006), others only from normative thresholds (AFNOR, 2008, 2004), and for certain system-specific-hazard the available vulnerability information is extremely limited. Without at least one vulnerability point linking a climate intensity to the start of damage (or to the fact that the damage may start) it is impossible to explicitly transform an intensity into a meaningful hazard for a given system.

The existing literature reflects this separation. Many so-called hazard maps are in fact return-intensity maps (Augusto Sanabria and Carril, 2018; Serra-Llobet et al., 2022; Sfetsos et al., 2023), representing climate extremes without explicitly incorporating system vulnerability. Other approaches rely on full vulnerability curves to compute damage from extreme intensities (Aznar-Siguan and Bresch, 2019; Mühlhofer et al., 2023), but in these cases the hazard remains defined as a physical intensity, and vulnerability intervenes only afterwards in the damage calculation. These approaches implicitly assume complete knowledge of vulnerability curves, which is not always realistic.

Thus there is currently no explicit, simple, and reproductible method to define a system specific hazard from a vulnerability point, in situations where vulnerability information is heterogeneous and may be limited to a single point. The present work addresses this methodological gap of a system-specific threshold and by operationalizing this definition through a model capable of producing truly system-specific hazard maps.

## 3. Concepts and Definitions

### 3.1. Hazard

Hazards are commonly defined as events or conditions that may cause damage to a given physical system. However, the link between hazard intensity and the system of interest damage is hard to determine.

The work of IPCC (2014) defines hazard as “The potential occurrence of a natural or human-induced physical event or trend or physical impact that may cause loss of life, injury, or other health impacts, as well as damage and loss to property, infrastructure, livelihoods, service provision, ecosystems, and environmental resources.” Emanuel (2011) and Eberenz et al. (2021) employed an intensity hazard threshold, which is different from zero, indicating the onset of loss for a given system of interest. In their hurricane focused studies, this threshold corresponds to wind speed. Westen and Greiving (2017) complement these views by identifying four technical characteristics of hazards: location, intensity, frequency, and probability.

Building on these foundations, we propose a system-specific definition of Climate Change-Related Hazard, placing the system at the core of hazard definition in the next section.

### 3.2. System-Specific hazard related to climate change

General definition: a climate change related hazard is the potential occurrence of a climate-related physical event that may cause damage to a specific system of interest.

This system-specific perspective can also consider internal heterogeneity: subsystems within an infrastructure can have distinct HITs. Thus, if needed, hazard modeling may consider subsystem-level vulnerability to capture potential damage. Although subsystem-level vulnerability modeling could improve accuracy, current hazard models rarely achieve system-specific hazard representation. This level of hazard definition is likely to be feasible only in highly specialized studies focused on critical infrastructures.

These definitions allow for refinement around the system of interest under study. A hazard intensity threshold (Dutel, 2025) signals the potential beginning of damage. Importantly, this definition distinguishes hazard data from general climate or meteorological data. When we speak about hazard data, it is in the light of a system.

### 3.3. Hazard Intensity Threshold

The HIT defines the transition from “No Hazard” to “Hazard”. HIT is the intensity level at which the PDR becomes greater than zero. Using vulnerability curves, HIT can also be set for specific damage levels. For example, $HIT_{PDR=0.2}$ corresponds to the intensity beyond which damage reaches 20%. This approach can be extended to other PDR values (e.g. 40%,60%,80%).

When a vulnerability curve is available, identifying HIT is straightforward: it is the point on the curve where damage begins or reaches a selected PDR. For instance, Nirandjan et al. (2024) provide vulnerability curves for various infrastructure types, enabling extraction of these thresholds.

Figure 2 the basic concept behind HIT derived from a vulnerability curve, and Figure 3 shows a $HIT_{PDR=0.2}$, corresponding to a 20% damage rate. Figure 4 presents the hazard modeling framework, which is part of the exposure modeling approach developed in Dutel et al. (2025).

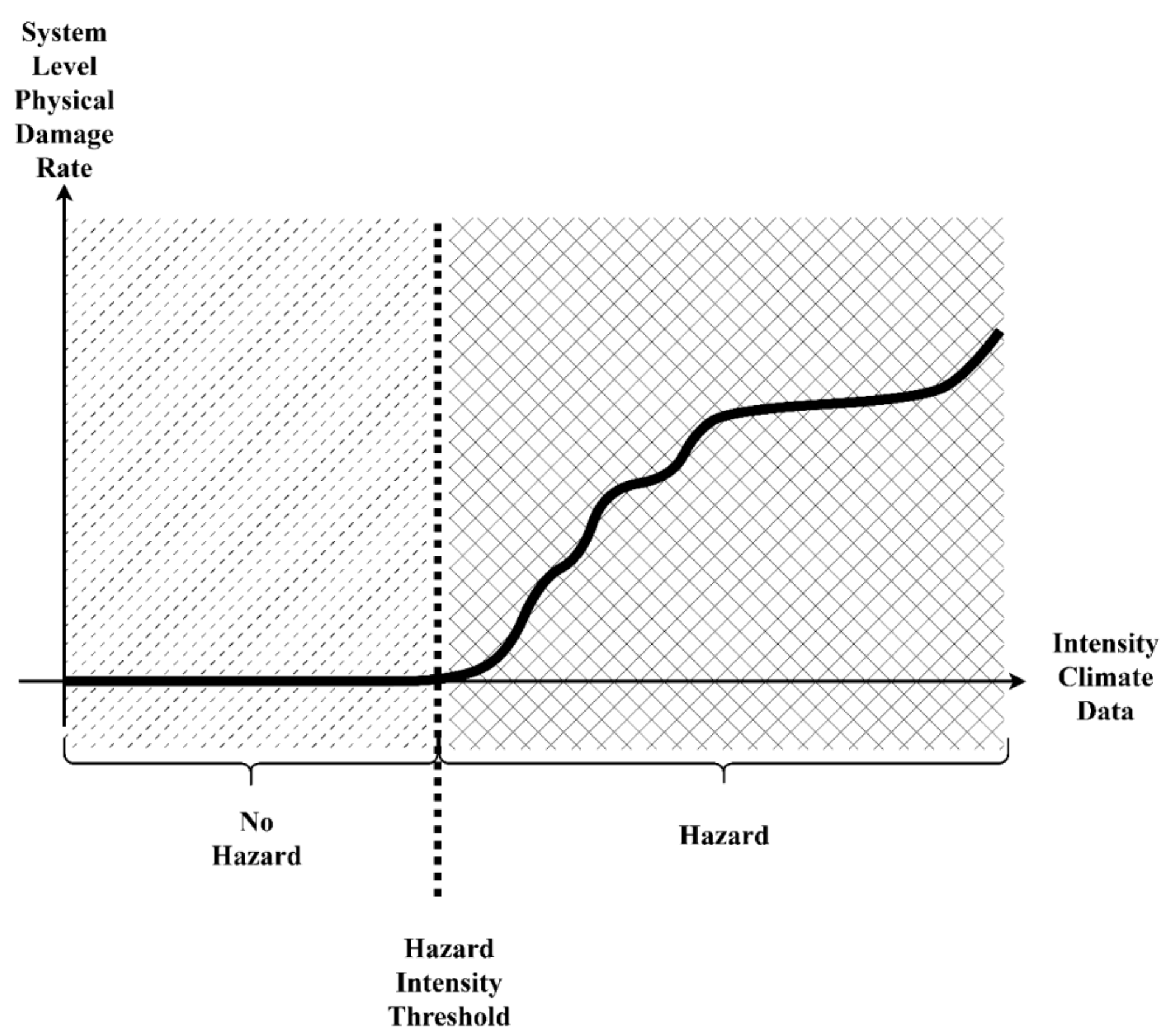

**Figure 2.** The Hazard Intensity Threshold was identified through a vulnerability curve.

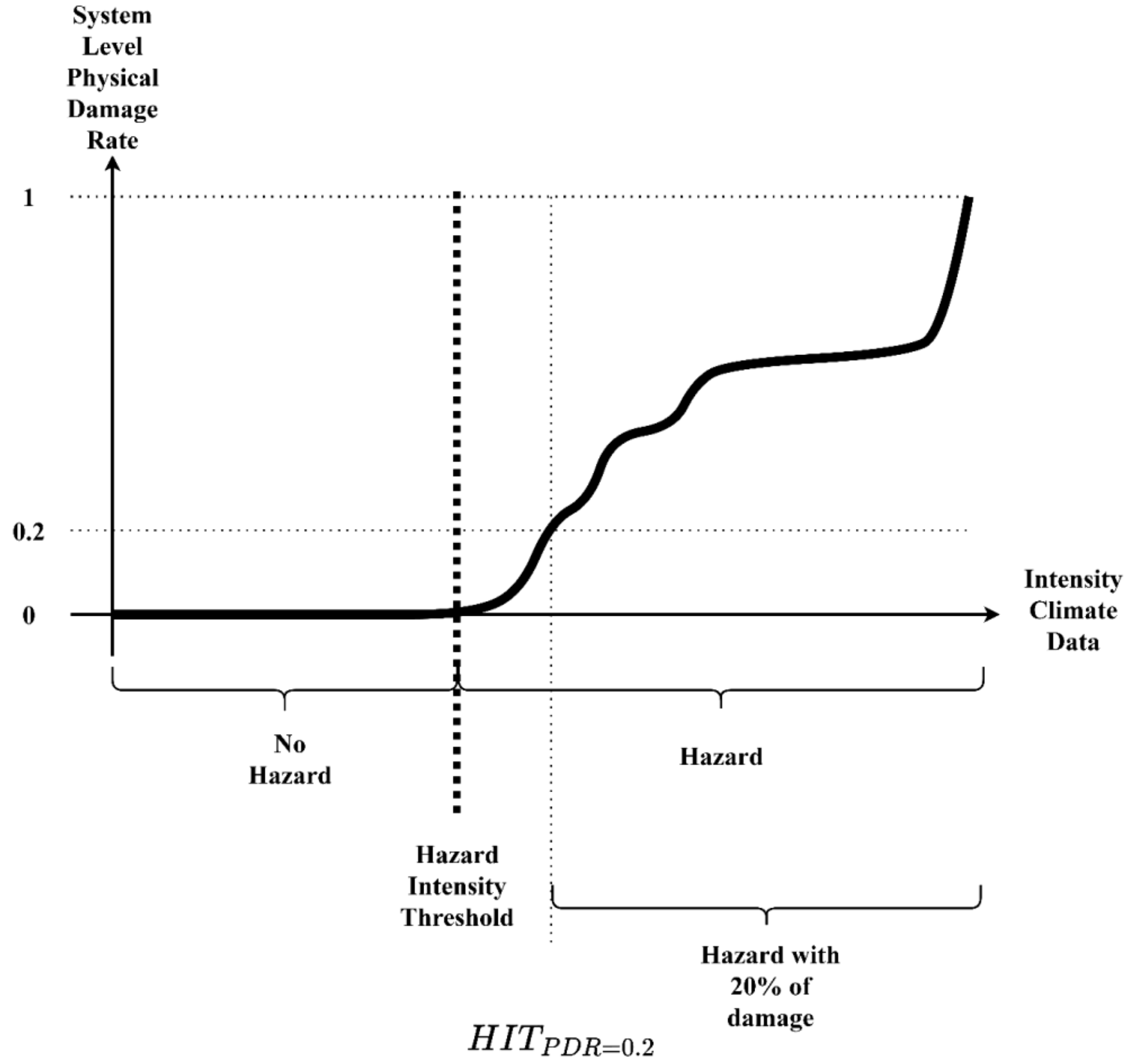

**Figure 3.** Figure to show a Hazard Intensity Threshold corresponding to 20%

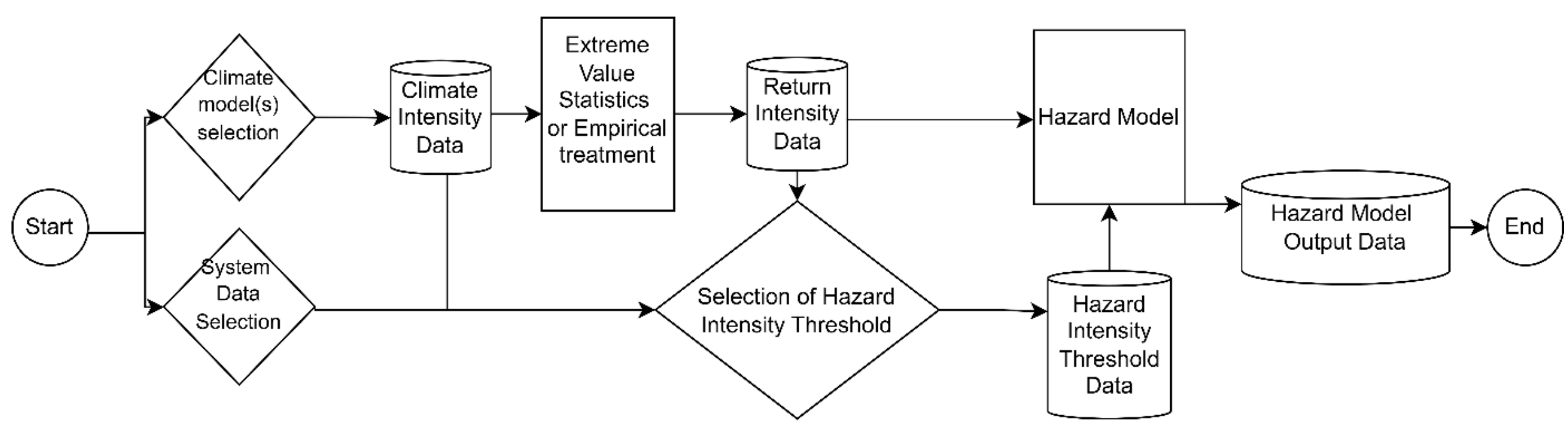

**Figure 4.** Hazard modeling framework. This framework is a subpart of the Exposure Modeling Framework developed in the work Dutel et al. (2025) .

Section 3 provides the concept, definitions, and illustrative figures that form the basis for applying the BHMP and for selecting HITs. These elements are essential for the modeling approach described in Section 4.

## 4. Models and Methods

Building on the concepts and definitions presented in Section 3, Section 4 introduces the modeling approach used in this study. We present the BHMP from Dutel et al. (2025), its formulation, and the methodology for determining HIT from vulnerability curves or design standards to generate system-specific hazard maps.

### 4.1. Boolean Hazard Model with Probability

The hazard model takes as input the return intensity data, the hazard intensity threshold data, and the corresponding spatial coordinates (longitude $x$, latitude $y$). The BHMP , introduced in Dutel et al. (2025), is formulated as:

$$H(i_T, x, y) = BHMP_{i_{HIT}, P_{i_{HIT}}}(i_T, x, y) = \begin{cases} 1, & if\ i_T > i_{HIT} + k_{^\circ C} \\ 0, & otherwise \end{cases} \quad (1)$$

Where $i_T\ (x, y)$ is the return intensity at location $(x, y)$, $i_{HIT}$ is the HIT at location $(x, y)$, $k_{^\circ C}$ is the optional constant (e.g., in °C) for multiclass thresholding. $P_{i_{HIT}}$ is the probability associated with the return period used to define the HIT.

This model enables system-specific hazard modelling based on intensity thresholds. The Section 4.2. is to determine these thresholds, which link climate intensities to the vulnerability of the system under study.

### 4.2. Obtaining the System Specific Hazard Intensity Threshold

Determining HITs is a critical step in system-specific hazard modeling because it links climate intensity to system vulnerability. A HIT represents the intensity level at which a system transitions from a state of no damage to a state where physical damage becomes possible. Beyond the basic HIT, HITs can also be defined for specific PDRs , such as 10% or 50%. Additionally, a constant $k_{^\circ C}$ can be added to $i_{HIT}$ to represent an adjusted threshold.

Methodology to obtain a HIT related to both climate data and an infrastructure type:

- Option 1: When a vulnerability curve is available for the infrastructure type, the intensity at which damage begins can be extracted directly. If the curve provides a continuous relationship between intensity and damage, thresholds for specific PDRs, such as 10%, can also be obtained.
  The PDR ($PDR$) represents the proportion of damage expected for a given system type and system level $s_j$ when exposed to an intensity of a hazard type $h$. It depends on both hazard intensity and system characteristics. Formally, for a system type $s_j$, for a hazard type $h$, with parameters $\theta(h, s_j)$ the vulnerability function is:

$$f_{\theta(h,s_j)}(i_h) = PDR_{s_j, i_h} \qquad (2)$$

  To determine the HIT for a given damage rate $\alpha$, invert the function:

$$i_{HIT_{PDR=\alpha}} = f_{\theta}^{-1}\left(PDR_{s_j, i_h} = \alpha\right) \qquad (3)$$

- Option 2: If no vulnerability curve is available, the HIT can be derived from design standards. These standards often specify the intensity levels that an infrastructure is expected to withstand under normal conditions. Identifying such design provides a practical way to link the infrastructure to the hazard type. Intensities thresholds in standards can be considered as point of vulnerability information of the concerned assets. However, these standards depend on the country and geographic zone, meaning that the thresholds may vary spatially and are not uniformly distributed. When the standard is respected, it indicates an intensity that the system can tolerate without damage.
- Option 3: If neither vulnerability curves nor design standards are available, and no design information exists for the infrastructure, additional data must be collected. Relevant options, not explored in this study, include the use of fragility curves. Otherwise, field inspections, engineering assessments, or numerical simulations can be performed to estimate the intensity levels that the system can withstand.

Here we will focus on option 1 and option 2.

The modeling framework directly addresses the research gap identified in this paper by providing system-specific hazard information, linking climate intensity data with vulnerability characteristics, and enabling hazard maps tailored to infrastructure types. Section 3 establishes the modelling approach and HIT methodology; we now turn to practical applications in Section 4.

## 5. Use Cases: Set Up & Results

This section demonstrates the generalization of the HIT to different contexts. In previous work Dutel et al. (2025), we explained that the methodology to set HITs is applicable to multiple hazard types and systems of interest. The strength of HIT lies in its ability to link climate data to the characteristics of a specific system. When integrated into the BHMP , HIT enables the generation of hazard information that is tailored to a particular physical infrastructure.

## 5.1. Use Case One: Hazard Model Outputs Based on a Generic Vulnerability Curve

To illustrate the methodology, we first apply the hazard modeling framework using a generic vulnerability curve proposed by Emanuel (2011). Although based on hypothesis and not fitted to empirical data, this curve provides a useful starting point for demonstrating the process of deriving HITs when vulnerability curve is available. Emanuel (2011) develops vulnerability curves for insured properties (mainly buildings) without distinguishing between building types or even distinguishing between insured properties. The curves express the PDR as a function of wind speed under two assumptions. From this work we use the vulnerability function that has a 50% damage at 203.7 $km.h^{-1}$ as shown in Figure 5.1.1.

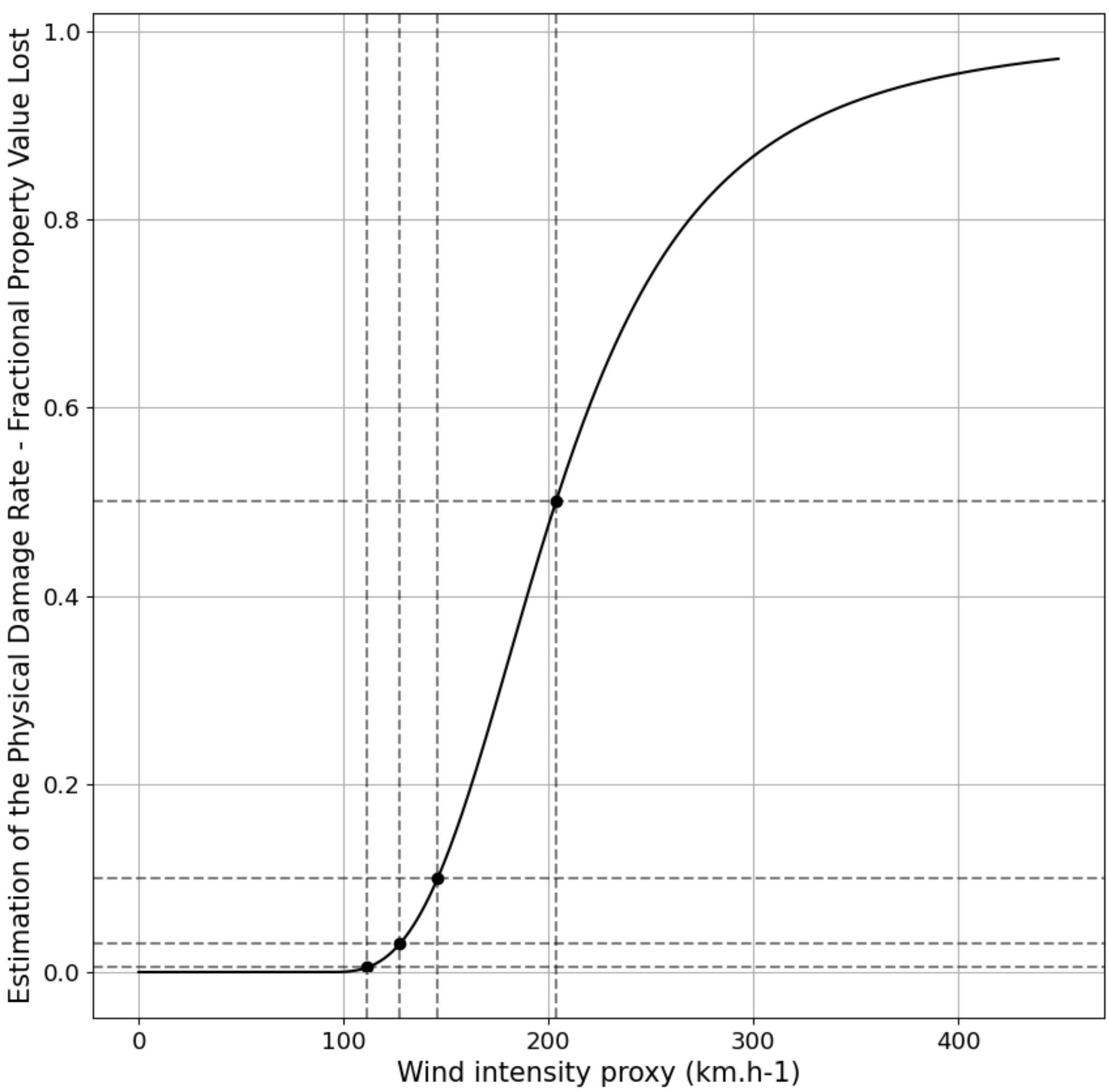

**Figure 5.** Vulnerability curve from Emanuel (2011). This hypothetical vulnerability curve represents the relationship between wind intensity (in $km.h^{-1}$) and the Physcial Damage Rate for insured properties. The curve is based on hypotheses rather than empirical data, with a start of damage at 92.6 $km.h^{-1}$ and a 50% damage at 203.7 $km.h^{-1}$. It serves as an example of a vulnerability function linking hazard intensity to expected damage for a given system.

Table 1 summarizes the corresponding wind speed thresholds.

| HIT corresponding to a PDR | Infrastructure Type: Insured Property – Mainly Buidings | Intensity $km.h^{-1}$ |
|---|---|---|
| HIT_0,5%=0,005 | | 111.6 |
| HIT_3%=0,03 | | 127.5 |
| HIT_10%=0,1 | | 146.0 |
| HIT_50%=0,5 | | 203.7 |

**Table.1.** Illustrative Table of Hazard Intensity Thresholds corresponding to different Physical Damage Rates (0.005,0.03,0.1,0.5) for one physical system type: insured property (mainly buildings). Note: The HIT values are derived from a hypothetical vulnerability curve proposed by Emanuel (2011) . We intend to show these thresholds solely as illustrative example for methodological purposes and does not represent an empirical or design-based threshold.

The HIT values were then used as inputs to the BHMP to generate hazard outputs for insured properties. It is important to note that this example combines past climate data for France with vulnerability information derived from insured properties in the United States. Consequently, the resulting maps should be interpreted as illustrative outputs of the methodology rather than operational hazard maps. This approach demonstrates the process and capabilities of the framework, while highlighting the need for system-specific vulnerability data to ensure accurate hazard modeling for practical applications in a context of climate change. Wind gust data we obtained from ERA5 reanalysis (1940-2024) post processed to compute empirical return intensities. Hazard maps were produced for four thresholds (111.6,127.5,146.0, and 203.7 $km.h^{-1}$), corresponding to the selected PDR levels.

Figure 6 presents the resulting hazard maps for an empirical return period of 42 years. The comparison between two time periods 1940-1982 and 1982-2024 reveals trends: hazard occurrences, defined as exceedances of HITs, are higher in the recent period 1982-2024 than in the earlier period 1940. No part of the territory exhibits a hazard corresponding to a PDR of 50% when using HIT values obtained from Emanuel (2011).

These results demonstrate the practical application of HIT obtained from vulnerability curves within the proposed methodology. Although this example is illustrative and based on generic vulnerability data, the resulting hazard maps are system-specific: insured properties; climate-data-specific: historical wind gusts from ERA5; and damage-level-specific: HIT corresponding to PDR thresholds. While this case relies on generic vulnerability curve, it illustrates the capability of the methodology to generate hazard maps that reflect both climate and system characteristics. In the following section, we extend this approach by using vulnerability curves that are specific to particular infrastructure types, thereby providing more accurate system-specific hazard information related to climate hazards.

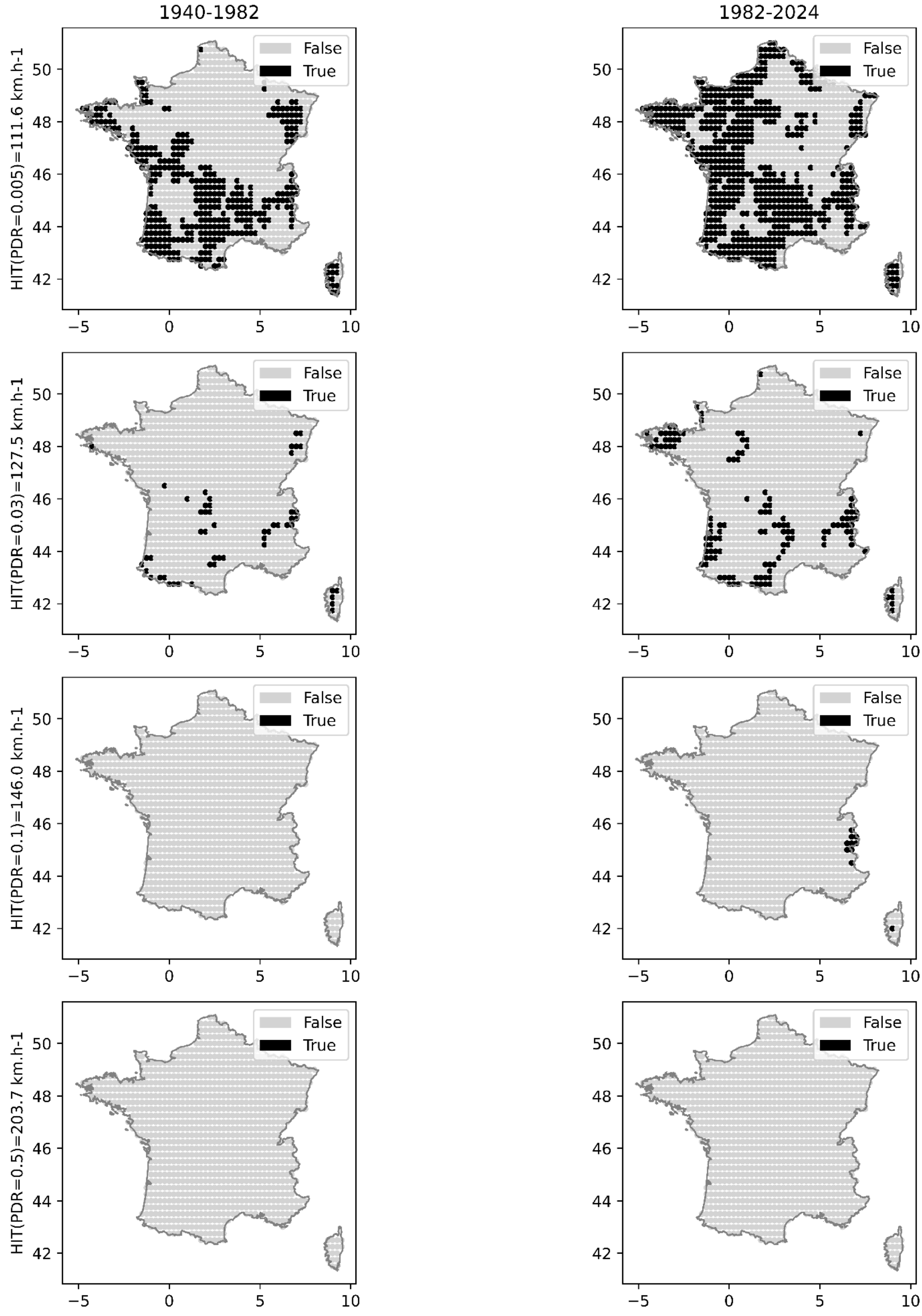

**Figure 6.** Boolean Hazard Model with Probability output map for Insured Property – Mainly Buildings for wind gust ($km.h^{-1}$) for an empirical return period of 42 years.

## 5.2. Use Case Two : Hazard Model outputs obtained thanks to HIT based on specific system vulnerability curve.

This use case illustrates the application of the proposed methodology to derive HITs from vulnerability curves for different infrastructure types. HITs are critical inputs for the BHMP , extracting infrastructure type specific threshold enable system-specific hazard mapping.

The vulnerability curves used in this illustrative case study are taken from Nirandjan et al. (2024) who compiled fragility and vulnerability curves for critical infrastructure systems across multiple hazards. For this use case, we focus on extreme wind with the same dataset as in the precedent use case ERA5 wind gust, we select two very precise infrastructure types:

- Wind Turbine (2.5 MW capacity, 80m hub height), from Jaimes et al., (2020) and collected by Nirandjan et al. (2024).
- Power Tower (design speed 120 $km.h^{-1}$), from Reinoso et al., (2020), and collected by Nirandjan et al. (2024).

These curves provide continuous relationship between wind intensity and PDR for each asset type.

The extraction process follows these four steps:

1. Select the PDR thresholds (e.g. 0.5%; 3%,10%,50%).
2. Identify corresponding wind intensities from the vulnerability curve.
3. Record HIT values for each infrastructure type and PDR level.
4. Use these HIT as inputs, together with the return intensities for hazard modeling with BHMP.

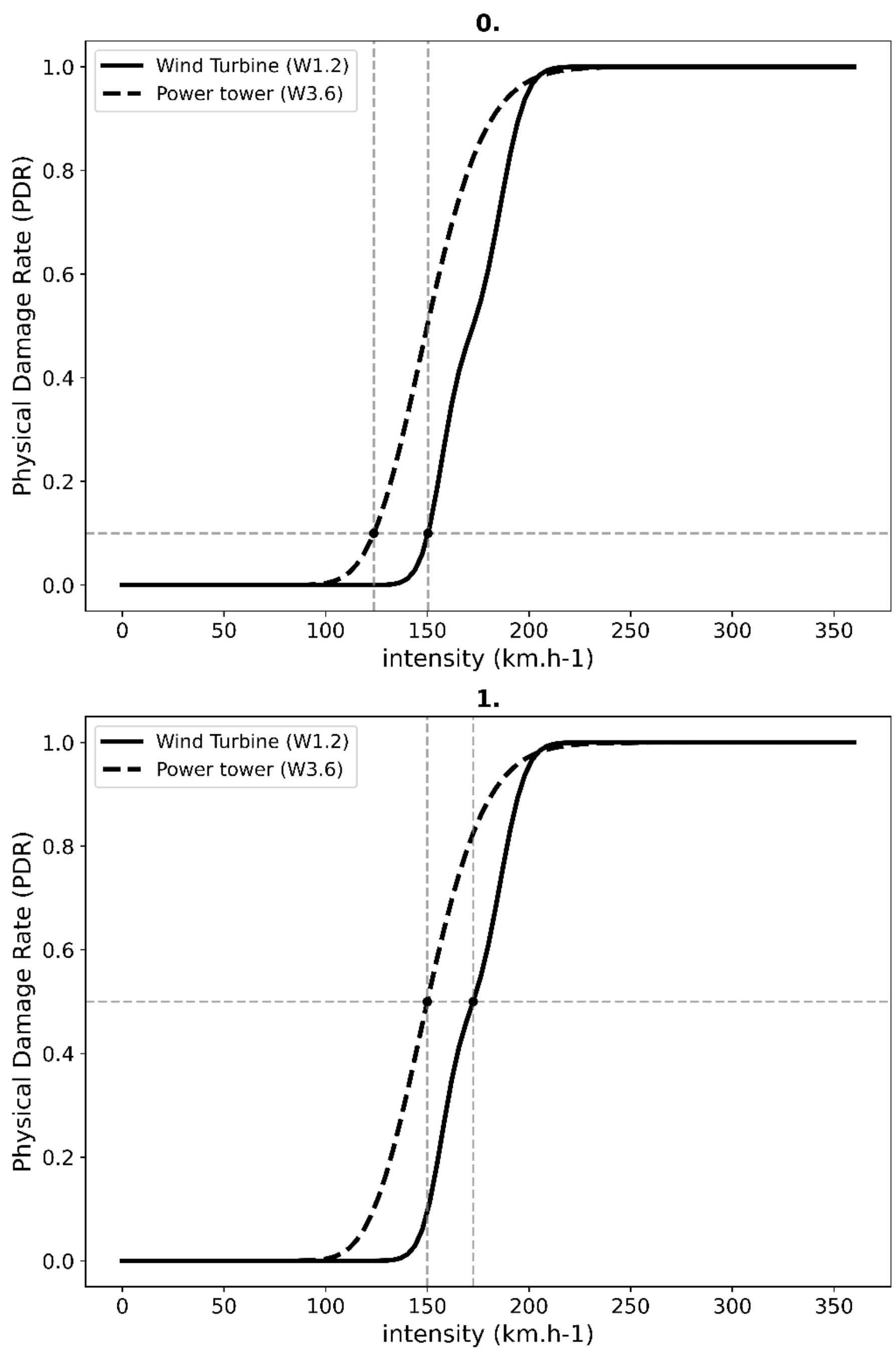

**Figure 7.** Vulnerability Curves collected by Nirandjan et al. (2024) to extract Hazard Intensity Threshold for Physical Damage Rate of 0.1 and 0.5. These vulnerability curves are for two different system types: Wind Turbine and Power Tower. Wind turbine : 2.5-MW capacity, 80-m hub height (Jaimes et al., 2020). Power tower: Design speed 120 $km.h^{-1}$ (Reinoso et al., 2020). Data source : Nirandjan et al., (2024). **0.** PDR = 0.1 and **1.** PDR=0.5.

Table 2. summarizes the HIT values obtained for the two infrastructure types.

| | System Type | Infrastructure | |
|---|---|---|---|
| | System Sub Type | Wind Turbine (2.5-MW capacity, 80-m hub height \| W1.2) | Power tower (Design speed 120 km/h \| W3.6) |
| | | Intensity $km.h^{-1}$ | Intensity $km.h^{-1}$ |
| HIT corresponding to a PDR | | | |
| HIT_0,5%=0,005 | | 136,8 | 101,7 |
| HIT_3%=0,03 | | 144,1 | 113 |
| HIT_10%=0,1 | | 150,3 | 123,7 |
| HIT_50%=0,5 | | 172,6 | 150 |

**Table 2. HIT values obtained for the infrastructure types:** Wind Turbine (2.5-MW capacity, 80-m hub height) and Power tower (Design speed 120 km/h) collected in Nirandjan et al. (2024).

We derive the thresholds specific to each system from the identified vulnerability points corresponding to (0.005, 0.03, 0.1, 0.5) physical damage rates, and the corresponding intensities. We then use these thresholds as inputs for the BHMP, along with the wind gust data. We then compute the results of BHMP for different time periods and different models. We thus have BHMP depending on a time period, on a threshold specific to climate data, here wind, and a specific system, here a specific wind turbine type, or a specific power tower type.

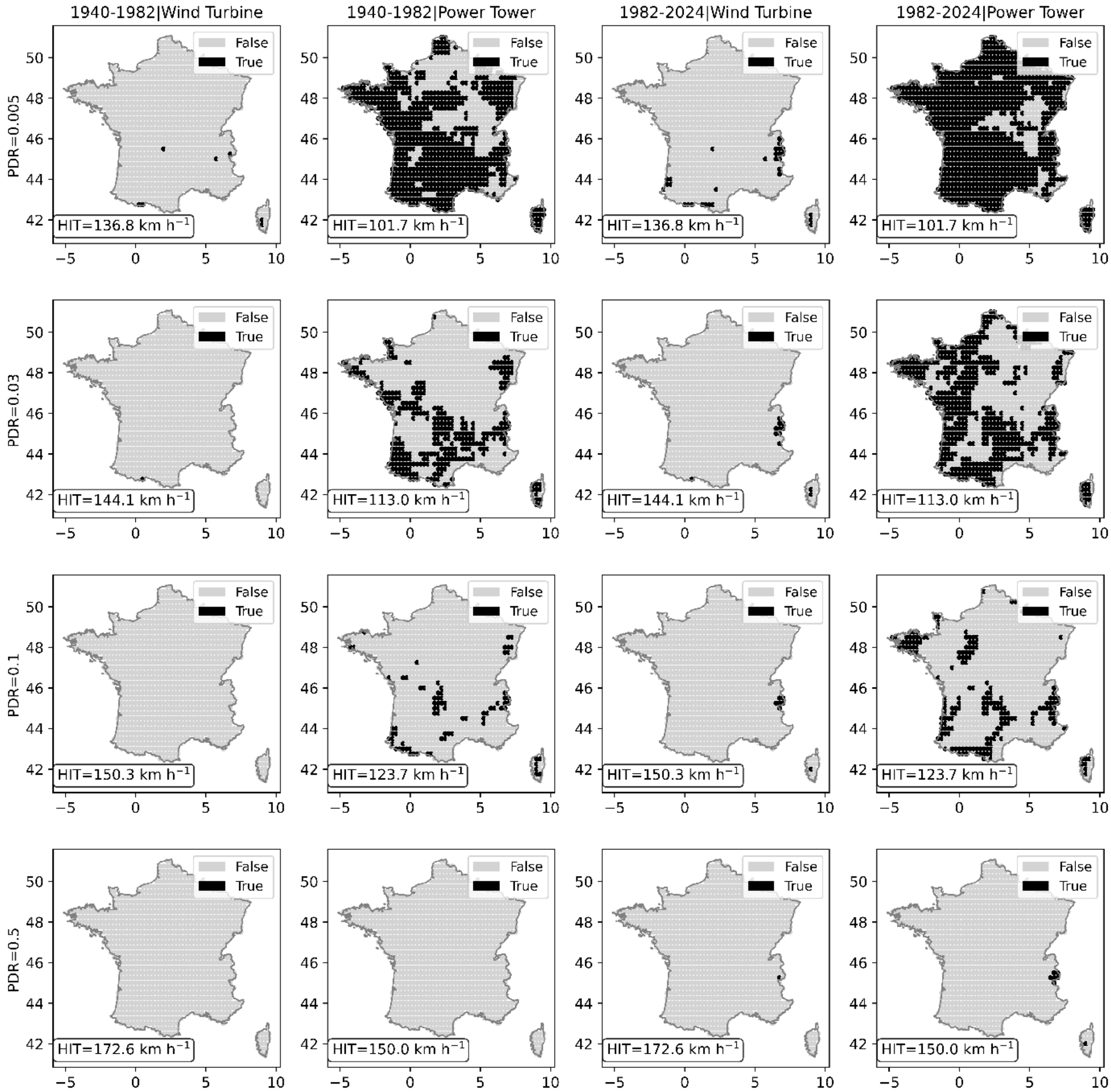

**Figure 8.** Boolean Hazard Model with Probability output maps for wind turbine and a power tower. Wind Turbine is a 2.5-MW capacity, 80-m hub height and Power Tower has a Design speed of 120 km/h.

For a given PDR, the spatial hazard footprints differ significantly between the wind turbine and power tower. The difference between these hazard maps, visible in Figure 8, reflects their distinct vulnerability curves: the power tower exhibits more hazard occurrences because its HIT values are lower (e.g., 101.7 $km.h^{-1}$ vs. 136.8 $km.h^{-1}$ for PDR=0.5%). These results are specific to the assumptions embedded in these vulnerability curves, systems designed under different assumptions or standards could exhibit different behaviors.

This use case illustrates a key point: a climate-related hazard, defined as the occurrence of a potential damage event for a system given the occurrence of a certain climate intensity, depends strongly on the system under study. An intensity that constitutes a hazard for one physical system may not be a hazard for another. This depends on the intensities taken into account in the design phase and the resulting characteristics of the system. In other words, the transition from no damage to the start of damage varies across infrastructure types, and infrastructure design hypotheses, even within the same broader system category (here, infrastructure of the electrical network; electrical system at a country scale: wind turbine and power tower). The hazard maps obtained here, which link BHMP outputs to specific damage levels, to specific systems with specific design hypotheses, are made possible by prior work on vulnerability that provides system-specific vulnerability curves and collection of system-specific vulnerability curves. These results provide actionable insights for exposure modeling, such as identifying locations where a given system should not be installed to avoid specific damage rates.

### 5.3. Use Case Three: Hazard Model outputs obtained thanks to Hazard Intensity Thresholds based on Design Standards

All this use case is methodological view of what was done in Dutel et al. (2025) for the hazard part.

This use case illustrates how normative design standards can serve as a source of HITs when detailed vulnerability curves are unavailable. Such an approach is particularly relevant for infrastructure types where regulatory design values are well established, as it enables hazard modeling aligned with current engineering practices.

We focus on structures which are concerned by the Eurocode (AFNOR, 2004, pp. 1991-1–5) and its French National Annex (AFNOR, 2008, pp. 1991-1–5). These standards define maximum air temperatures under shelter with an annual exceedance probability of 0.02 (i.e. return period of 50 years), which we adopt as HIT values. Figure 4.3.1. shows the spatial distribution of these thresholds across French departments, ranging from 35°C to 40°C.

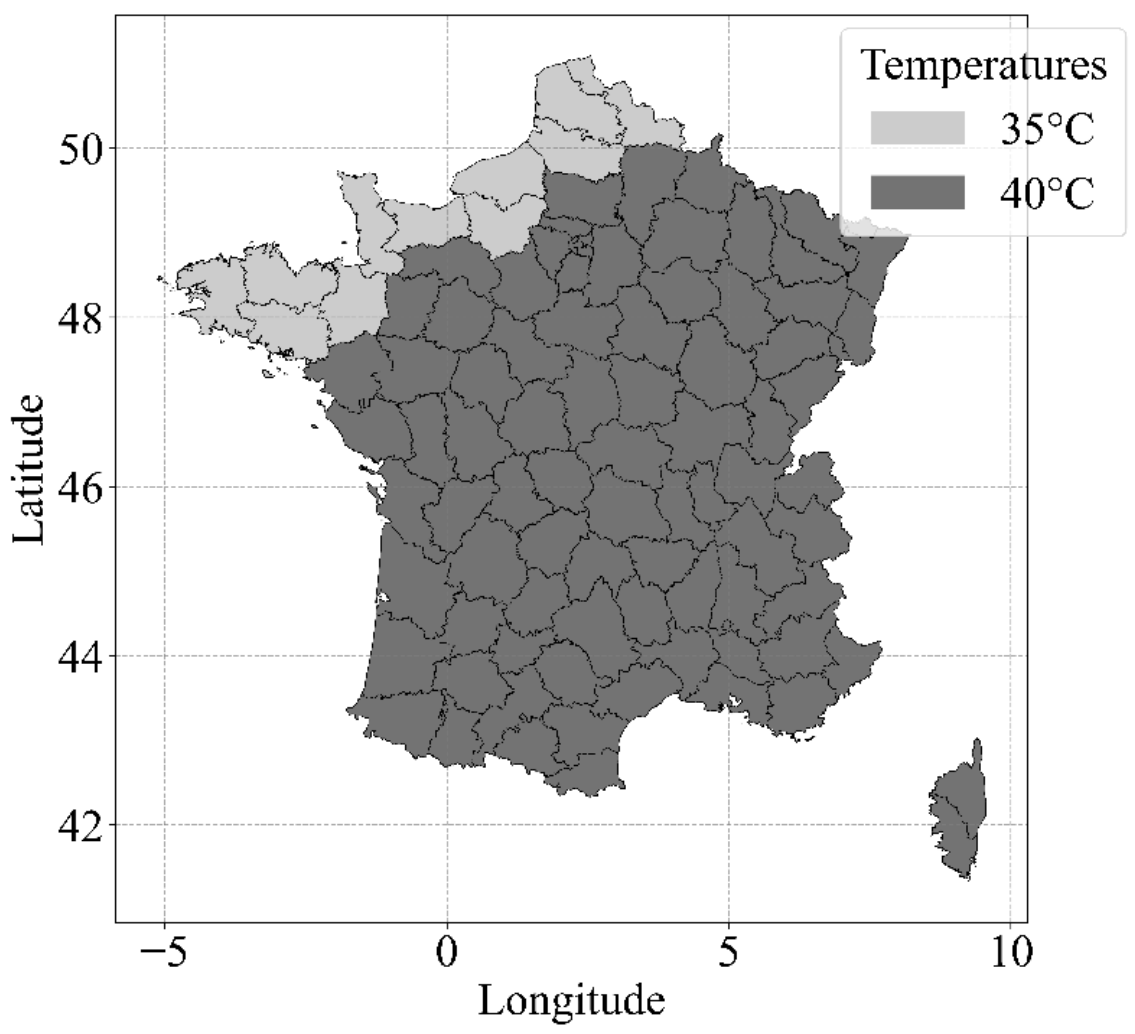

**Figure 9.** Maximum air temperature under shelter, by French department from Clause 6.1.3.2 (1) of NF EN 1991-1-5/NA. Figure from Dutel et al. (2025).

The extraction process involves:

1. Identifying the relevant design standard for the infrastructure type.
2. Extracting the normative return intensity for each location.
3. Using these HIT values as inputs to the BHMP.

Applying BHMP with Eurocode-based HITs produces hazard maps that shows the exceedance of the design assumptions with return intensities that are lower or equal to the return intensity of the Eurocode, and here also only with the empirical return intensities. These hazard maps highlight locations where the design threshold is exceeded. It is input information for exposure modeling. Unlike vulnerability-curve-based HITs, which allow multi-level damage modeling, normative HITs provide a binary perspective: compliance versus exceedance. And the exceedance of the HIT which is based on threshold does not mean that there is occurrence of a damage but that a damage can happen because of the exceedance of the threshold that was taken in the design of the infrastructure. The produced hazard maps are really interesting they are showing the evolution of the hazard footprint for a given system type, seeing in which location the intensity is constituting a hazard.

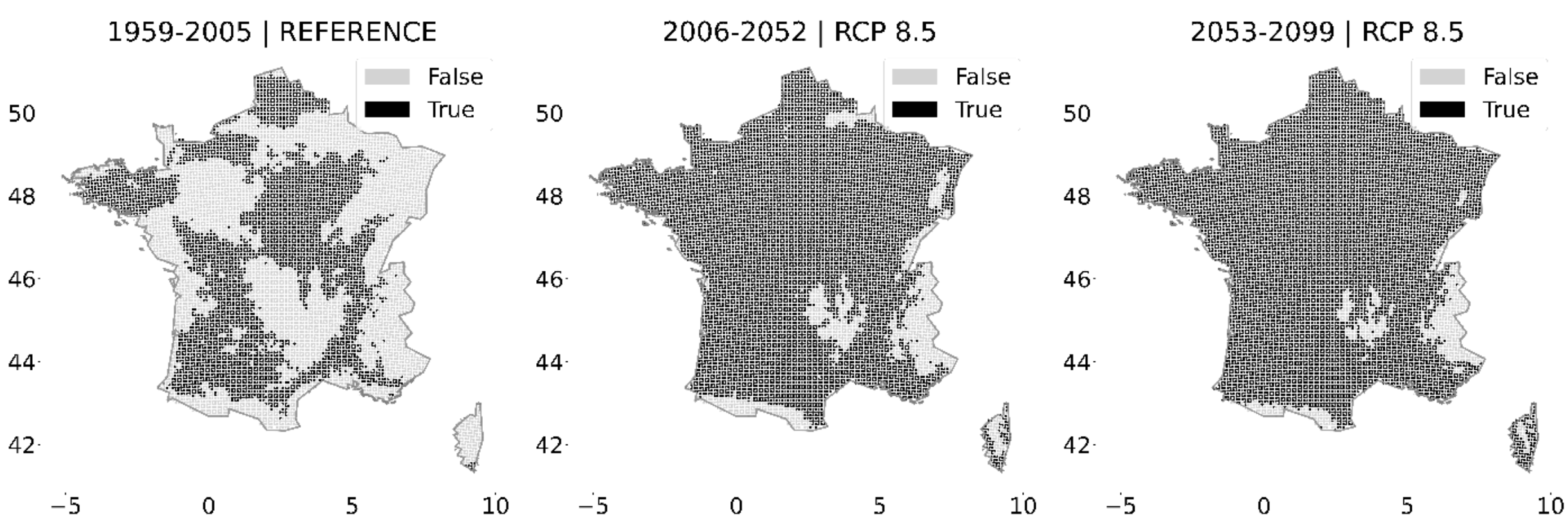

**Figure 10**. Hazard map based HIT based on standard: Eurocode (AFNOR, 2004, pp. 1991-1–5) and the associated National Annex (AFNOR, 2008, pp. 1991-1–5) this precise method is used in Dutel et al. (2025).

## 5.4. Hazard Modeling Flexibility Across Vulnerability Data Sources

The three use cases illustrate the generalization of the BHMP across multiple hazards and system definition. BHMP, previously applied to extreme temperature hazards with Eurocode based thresholds, is here extended to extreme wind and applied to systems ranging from broadly defined insured properties to precisely defined infrastructures such as wind turbines and power towers. This demonstrates BHMP's adaptability to different intensity metrics, hazard types, and sources of vulnerability information.

Beyond hazard diversity, the use cases reveal the fact that HIT can adapt to different systems levels. Thus, the hazard maps can be useful for different system levels. The first case applies a generic vulnerability curve to insured properties, enabling hazard maps for multiple PDRs but with moderate confidence in thresholds due to underlying assumptions. The second case focuses on two precisely defined infrastructure types supported by detailed vulnerability curves, allowing HIT extraction for several PDR levels, the fact that HITs are very system specific increase the confidence. The third case uses normative Eurocode design standards to define HITs for a large set of systems. These thresholds are highly reliable because they are based on regulatory design assumptions, but their interpretation requires caution: the exceedance of such a HIT does not imply that damage has occurred; rather, it indicates that damage becomes possible because the intensity has surpassed the level considered during design.

## 6. Discussion

The use cases presented in this study collectively highlight the flexibility of the BHMP. Unlike previous work, where BHMP was used solely as an input for exposure modeling, this paper positions BHMP as a standalone process. The model is generalized to handle multiple hazard types- extreme temperature (threshold expressed in °C) and extreme wind (threshold expressed in $km.h^{-1}$)- while maintaining a consistent logic.

Beyond hazard diversity, the use cases reveal different levels of system definition and confidence in HITs . The first case applies a generic vulnerability curve to insured properties, enabling hazard maps for multiple PDR . However, this vulnerability curve is highly generic, making it difficult to apply to very specific systems or lower system levels. Nevertheless, it demonstrates the use of vulnerability curves to derive HITs. The main limitation here is the lack of precision: finer information for specific systems is needed to improve hazard model instantiation and outputs, as well as the specificity of system-specific hazard modeling in a climate change context. This is addressed in use case 2.

The second use case focuses on two precisely defined infrastructure types supported by detailed vulnerability curves. This allows HIT extraction for several PDR levels with higher confidence compared to use case one.

The third use case uses normative Eurocode design standards to define HITs for a large set of systems. These thresholds are reliable due to their basis in regulatory design assumptions. However, their interpretation requires caution: the exceedance of a HIT does not imply that damage has occurred; rather, it indicates that damage may occur because the intensity has surpassed the design threshold. This distinction is essential for considering the maps of use case 3, which illustrate the evolution of hazard footprints for a given system type and identify locations where intensities constitute a hazard under current design assumptions.

Together, these results demonstrate BHMP's capacity to produce hazard maps that reflect both climate intensity and at least one point of information about system's vulnerability, offering actionable insights for infrastructure design and adaptation planning. This approach introduces multi-threshold hazard maps based on PDR levels, comparative analysis of HIT sources, and visualization of system specific hazard footprint evolution- key contributions for risk-informed decision-making in the context of climate change.

It is important to note that for the first two use cases (extreme wind), only historical ERA5 data was used, separated into distinct periods to analyze temporal changes. The return intensities are empirical and do not include climate projections. Incorporating wind gust projections would be a valuable extension for future work. In contrast, the third use case focuses on extreme temperature intensities using Eurocode-based thresholds and a temperature database that includes future climate projections. This demonstrates that BHMP is adaptable to different types of data: historical only datasets and datasets incorporating climate projections. This adaptability is critical in the context of climate change, as understanding the spatial distribution of future extremes for system-specific hazards become essential for adaptation planning. Such hazard outputs maps can support decisions for existing physical systems, inform the planning of future infrastructures, and improve knowledge about the evolution of climate related hazards. Furthermore BHMP is flexible to incorporate diverse sources of vulnerability information, multiple hazard types, and evolving climate datasets, making it a scalable framework for future multi-hazard and multi-infrastructure assessments.

The approach developed in this study focuses on generating hazard maps that are system specific and grounded in empirical return intensity data. We put the focus, in this paper on the hazard intensity threshold and the hazard modeling process, we use empirical return intensity not to add statistical modeling and in order to focus solely on HIT and system specific climate-change hazard definition and hazard modeling. This choice allows us to concentrate on the core question: how to obtain hazard outputs that accurately reflect the characteristics of the system under study.

The results show that hazard maps can be tailored to specific systems by carefully selecting HIT values based on available vulnerability information. Two sources of information were explored: vulnerability curves and design standards. Nevertheless, other sources of information could be explored in future works, for example fragility curves to fix hazard intensity thresholds. Vulnerability curves provide threshold for both the start of damage and different physical damage rates, while design standards indicate a vulnerability point where there is the supposition of not having damage at this point (but maybe it is  not the last point not implying damage, but from this point there is an uncertainty about having a damage or not, except if a vulnerability curve exist for this specific system). These sources enable the creation of hazard maps that are not generic but adapted to the system's physical characteristics.

This system-specific approach ensures that the produced hazard maps are meaningful for a given infrastructure. These hazard outputs can serve two purposes: (i) to provide hazard information as it is, in order to see system-specific hazard output spatial repartition (ii) as precise inputs for exposure modeling frameworks. By applying the Boolean Hazard Modeling with Probability to different illustrative use cases, we show that it is applicable to different systems, hazard types and different sources of vulnerability information (standard, vulnerability curve).

## 7. Conclusion

In this study, we focused on producing hazard maps that are system-specific and grounded in existing data. The approach relies on return intensity values, here calculated empirically, combined with carefully selected HITs . This choice allowed us to concentrate on two key aspects: (i) how to obtain HIT values from available sources of vulnerability information, and (ii) how to generate hazard maps tailored to specific systems or infrastructure types.

The results demonstrate that system-specific hazard maps can be constructed using different sources of vulnerability information, such as vulnerability curves and design standards. Vulnerability curves provide thresholds that reflect either the start of damage or specific damage rates. For design standards, the threshold usually marks a point where no damage should occur. Beyond this point, damage becomes possible, and the event can be classified as a hazard. In both cases, the HIT ensures that hazard modeling captures the characteristics of the system under study.

By applying this approach across different systems and hazard types, we show that hazard maps can serve as precise inputs for exposure modelling.

Future research should aim to explore additional source of vulnerability data to find relevant HIT to other systems, subsystems and hazards, and thus compute the corresponding hazards.

**CRediT authorship contribution statement**

Matthieu Dutel: Conceptualization, Methodology & Models, Software, Validation, Formal Analysis, Investigation, Data Curation, Writing-Original Draft, Visualization.

**Acknowledgement**

Matthieu Dutel thanks Sixense and Sixense' clients, engineers, asset managers and experts; thanks, LGI team for research advice; thanks Jean Meunier Pion for scientific listening. Thanks Anne Barros, Adam Abdin, Didier Soto for the supervision of the thesis. Authors acknowledge that this paper is made in the context of a CIFRE thesis with CentraleSupélec and Resallience by Sixense Engineering, VINCI group until 28.02.2026 and personally financed by M. Dutel. from 28.02.2026 to 30.05.2026.

**Références**